\documentclass[aps,preprint,nofootinbib,preprintnumbers]{revtex4}%Big
\usepackage{graphicx,amsmath,amssymb,bm,subfigure,comment}
\usepackage[english]{babel}
 \usepackage{color}
\usepackage{slashed}
\usepackage{amsmath}
\usepackage{wasysym}
\usepackage{latexsym}
\usepackage{graphicx}
\usepackage{amssymb}
\usepackage{amsmath}
\usepackage{mathrsfs}
\usepackage{feynmp}
\usepackage{fancyhdr}
\usepackage{color}

\newcommand{\beq}{\begin{equation}}
\newcommand{\eeq}{\end{equation}}

\begin{document}
 \title{Below the Breitenlohner-Freedman bound \\ in the nonrelativistic AdS/CFT correspondence}
 \author{Sergej Moroz}
 \affiliation{Institut f\"{u}r Theoretische Physik Universit\"at Heidelberg Philosophenweg 16, D-69120 Heidelberg, Germany}

 \begin{abstract}
 We propose that there is no analogue  of the Breitenlohner-Freedman stability bound on the mass of a scalar field in the context of the nonrelativistic AdS/CFT correspondence. Our treatment is based on an equivalence between the field equation of a complex scalar in the AdS/CFT correspondence and the one-dimensional Schr\"odinger equation with an inverse square potential. We compute the two-point boundary correlation function for $m^2<m_{BF}^{2}$ and discuss its relation to renormalization group limit cycles and the Efimov effect in quantum mechanics.  The equivalence also helps to elucidate holographic renormalization group flows and calculations in the global coordinates for Schr\"odinger spacetime.
 \end{abstract}
 
% \pacs{}
 
 \maketitle
\section{Introduction}
The anti-de Sitter/ conformal field theory (AdS/CFT) correspondence \cite{AdSCFT} is a powerful technique which allows us to investigate gauge theories and develop some intuition about their behavior at strong coupling. Recently, Son \cite{Son} and Balasubramanian and McGreevy \cite{McGreevy} extended the technique to the realm of nonrelativistic physics.\footnote{A related, but Galilean noninvariant version of the nonrelativistic holography was constructed in \cite{Kachru}.} Their work was mainly motivated by the rapid progress and strong interest in the theory of cold fermions at unitarity \cite{Bloch, giorgini} which is strongly coupled and described by the effective nonrelativistic conformal field theory \cite{Nishida}. The proposal in \cite{Son, McGreevy} stimulated a considerable theoretical progress and led to a number of interesting insights (for review see \cite{Hartnoll}). %(a very incomplete list of original references is \cite{nonrelativistic}, see also a review \cite{Hartnoll}).

It often happens in physics that two apparently different physical problems have the same solution because they are described by the same mathematical equations. In this case it may be helpful for a better understanding of one of the problems to reformulate it in the language of the other one. In this paper we use one example of this equivalence, also recently mentioned in \cite{FB, Kaplan}, between the field equation of a complex scalar in the anti-de Sitter (or Schr\"odinger) background spacetime and the one-dimensional Schr\"odinger equation with an inverse square interaction potential defined on the  real positive half-line. We argue that contrary to the known presence of a stability mass bound in the Minkowski and anti-de Sitter spacetimes, there is no restriction on the mass of a scalar field in the nonrelativistic holography. We arrive to this conclusion by performing calculations in the Poincar\'e and global coordinates. Additionally, the quantum mechanical analogy allows us to gain a simple understanding of the regime, where a single bulk theory describes two different conformal field theories on the boundary \cite{Dobrev, Klebanov}. We construct renormalization group flows between the two CFTs and find the quantum mechanical interpretation of the (in the nonrelativistic case spurious) Breitenlohner-Freedman (BF) mass bound $m_{BF}^2$ \cite{BF}. Finally, using a standard AdS/CFT prescription, we compute the scalar two-point correlation function in momentum space for $m^2<m_{BF}^2$ in the nonrelatvistic AdS/CFT, discuss its properties and comment on the connection to renormalization group limit cycles and the well-known Efimov effect in quantum mechanics. In two quantum mechanical problems, we also propose examples of local composite operators which might be dual to the scalar field with $m^{2}<m_{BF}^{2}$ in the framework of AdS/CFT.

\section{Mass stability bound in $Mink_{d}$, $AdS_{d+1}$ and $Sch_{D+3}$ spacetimes} \label{instability}
In this section we identify a stability bound on the mass  of a free scalar field in different background spacetimes.
As a warm-up we consider a free complex\footnote{The mass stability bounds presented in this section for complex fields are valid also for real scalars in $Mink_d$ and $AdS_{d+1}$ spacetimes. However, in the nonrelativistic version of AdS/CFT one must necessarily use complex fields in $Sch_{D+3}$ spacetime to describe massive nonrelativistic particles in the boundary field theory.} scalar field $\phi$ of mass $m$ in the Minkowski spacetime $Mink_d$, defined by the action
\beq \label{ins1}
S[\phi, \phi^{*}]=-\int d^dx \left( \eta^{\mu\nu}\partial_{\mu}\phi^*\partial_{\nu}\phi+m^2\phi^{*}\phi\right),
\eeq
where the metric signature convention is $\eta_{\mu\nu}=\text{Diag}(-1,1,1,...)$.
The simplest solutions of the field equation $\Box\phi-m^2\phi=0$ are the plane waves
\beq \label{ins2}
\phi^{(\text{p-w})}=\exp[-iq^0 t+i \vec{q}\cdot \vec{x}],
\eeq
where $q^0$ is energy, $\vec{q}$ is momentum and $q^2\equiv-(q^{0})^2+\vec{q}^2$. The energy and momentum must satisfy the on-shell condition $q^2=-m^2$. We note that for $m^2<0$ we obtain a tachyonic solution, which is unstable. This is because in the range $\vec{q}^2<|m^2|$ the energy $q^0$ is pure imaginary and the solution $\phi^{(\text{p-w})}$ can grow exponentially in time. The instability also manifests itself in the energy-momentum tensor which for the free complex scalar $\phi$ is generally given by\footnote{There is an ambiguity in the definition of the energy-momentum tensor $T_{\mu\nu}$ reflected in the presence of the last term in Eq. (\ref{ins2a}). This contribution originates from the coupling of the scalar field $\phi$ to the scalar curvature $R$ of the background spacetime,  i.e. $\sim \chi \int d^d x \sqrt{-g} R |\phi|^2$. For the detailed discussion of the anti-de Sitter spacetime case see \cite{BF}.}
\beq \label{ins2a}
T_{\mu\nu}=[\partial_{\mu}\phi^{*}\partial_{\nu}\phi+\partial_{\nu}\phi^{*}\partial_{\mu}\phi]-g_{\mu\nu}[\partial\phi^{*}\cdot\partial\phi+m^2|\phi|^2]+\chi[g_{\mu\nu}\Box-D_{\mu}\partial_{\nu}+R_{\mu\nu}]|\phi|^2,
\eeq
where the metric $g_{\mu\nu}=\eta_{\mu\nu}$, the covariant derivative $D_{\mu}=\partial_{\mu}$, and the Ricci tensor $R_{\mu\nu}=0$ in the Minkowksi case. The energy-momentum tensor (\ref{ins2a}) is conserved for an arbitrary value of $\chi$. For the plane-wave solution (\ref{ins2}) the energy-momentum tensor simplifies
\beq \label{ins2b}
T^{(\text{p-w})}_{\mu\nu}=2q_{\mu}q_{\nu}.
\eeq
The energy density $\epsilon^{(\text{p-w})}=-(T^{(\text{p-w})})^{0}_{0}=T^{(\text{p-w})}_{00}$ of the tachyonic plane wave is negative for $\vec{q}^2<q^2$ suggesting the presence of the instability. In general, a field theory with $m^2<0$ in $Mink_{d}$ is stabilized by an addition of repulsive interactions. Specifically, in the quantum field theory an effective potential becomes bounded from below and the condensate $\langle \phi \rangle\ne 0$ is formed. This defines a new vacuum, around which solutions of the theory must be  expanded.

In the context of the AdS/CFT correspondence \cite{AdSCFT} consider  a free complex scalar in the anti-de Sitter spacetime $AdS_{d+1}$ with the action
\beq \label{ins3}
S[\phi, \phi^{*}]=-\int dz d^{d}x\sqrt{-g} \left( g^{\mu\nu}\partial_{\mu}\phi^*\partial_{\nu}\phi+m^2\phi^{*}\phi \right),
\eeq
where the $AdS_{d+1}$ metric in the Poincar\'e patch is given by\footnote{Henceforth we take the radius of the $AdS$ spacetime to be $\mathscr{R}=1$.}
\beq \label{ins3a}
ds^2=\frac{dz^2+\eta^{\mu\nu}dx_{\mu}dx_{\nu}}{z^2}
\eeq
with $z\in[0,\infty)$ denoting the radial $AdS$ coordinate.
If we perform a Fourier transform to the momentum space $x^{\mu}\to q^{\mu}$ on the boundary, the field equation can be written as
\beq \label{ins4}
\partial_{z}^{2}\phi-\frac{d-1}{z}\partial_{z}\phi-\frac{m^2}{z^2}\phi-q^2\phi=0, \qquad q^2=-(q^{0})^2+\vec{q}^2,
\eeq
which after the change of variables $\phi=z^{(d-1)/2}\psi$ can be expressed as
\beq \label{ins5}
-\partial_{z}^{2}\psi+\frac{m^2+\frac{d^2-1}{4}}{z^2}\psi=-q^2\psi.
\eeq
This is a one-dimensional Schr\"odinger equation,\footnote{In our convention $\hbar=2\mathscr{M}=1$ with the particle mass $\mathscr{M}$.} defined on the real positive half-line, with the classically scale invariant inverse square potential of the strength $\kappa=-m^2-\frac{d^2-1}{4}$ and the energy $E=-q^2$. This quantum mechanical problem was studied extensively \cite{Case,Furlan,Kolomeisky,Gupta,Beane:2000wh,Barford:2002je,Bawin:2003dm, Camblong, Ho,Braaten,Barford:2004fz,Hammer:2005sa,Hammer:2008ra} and by now is well understood: The inverse square potential is on the boundary between regular and singular potentials \cite{FLS} and must be regularized near the origin $z=0$. Depending on the value of the coupling constant $\kappa$, the solution of the Schr\"odinger equation has two qualitatively different regimes.  While for $\kappa<\kappa_{cr}=\frac{1}{4}$ there are no bound states and the energy spectrum is continuous and positive, for $\kappa>\kappa_{cr}$ an infinite bound state spectrum develops. The bound state spectrum is geometric, i.e. the ratios of energies of the adjacent levels are constant, with the accumulation point at $E=0$. In analogy with the Minkowski case the instability in Eq. (\ref{ins4}) can appear only when $q^2>0$. In the quantum mechanical language this corresponds to $E<0$. For instability actually to appear we additionally require that the corresponding bound state wave function $\psi=z^{(1-d)/2}\phi$ is physically acceptable due to Calogero \cite{Calogero}, i.e. that both $|\psi|^2$ and $\psi\partial_z \psi$ are continuous functions regular at the origin. Both conditions for instability are fulfilled only for $\kappa>\kappa_{cr}$. From Eq. (\ref{ins5}) this gives rise to a bound on the possible mass $m^2$ of a scalar in $AdS_{d+1}$
\beq \label{ins6}
m^{2}\ge m_{BF}^2=-\frac{d^2}{4},
\eeq
which is the BF bound in the anti-de Sitter spacetime. It was first derived in \cite{BF} by demanding positivity of the conserved energy functional for scalar fluctuations which vanish sufficiently fast at spatial infinity. Below the BF bound the $AdS_{d+1}$ background becomes unstable. To stabilize the theory bulk interactions must be introduced, which deform the AdS metric. This often leads to a formation of an IR wall \cite{Freedman} at some $z=z_{IR}$ (see also \cite{Kaplan} for a simple realization of this kind of deformation). In the boundary theory the IR momentum scale $\Lambda_{IR}=z_{IR}^{-1}$ is dynamically generated and the boundary operator $O$ dual to the bulk field $\phi$ acquires a nonzero expectation value even in the absence of an external source $J$. 

Recently, the concept of holography was extended to nonrelativistic physics \cite{Son, McGreevy}. The key idea is to investigate Einstein gravity (and its extensions) on the Schr\"odinger spacetime $Sch_{D+3}$ background with the metric in the Poincar\'e coordinates given by
\beq \label{ins7}
 ds^2 = -\frac{dt^2}{z^4} + \frac{-2 dt d\xi + dx^idx^i + dz^2}{z^2}, \qquad i=1,...,D.
\eeq
The isometries of the metric (\ref{ins7}) form the so called Schr\"odinger group \cite{Sch}. The dual nonrelativistic field theory is defined on the $D+2$-dimensional anisotropic conformal boundary \cite{Horava} with the metric
\beq
\widetilde{ds^2}=-dt^2-2dtd\xi+dx^i dx^i.
\eeq
Consider a free complex scalar of mass $m_0$ on the $Sch_{D+3}$ background defined by
\beq \label{ins8}
S[\phi, \phi^{*}]=-\int dz dt d\xi d^{D}x\sqrt{-g} \left( g^{\mu\nu}\partial_{\mu}\phi^*\partial_{\nu}\phi+m_{0}^2\phi^{*}\phi \right).
\eeq
After transforming to the momentum space $(t,\xi,\vec{x})\to(\omega,M,\vec{q})$, the field equation reads \cite{Son}
\beq \label{ins9}
\partial_{z}^{2}\phi-\frac{D+1}{z}\partial_{z}\phi-\frac{m^2}{z^2}\phi-\widetilde{q}^2\phi=0, \qquad \widetilde{q}^2\equiv-2M\omega+\vec{q}^2,
\eeq
with $m^2=m^2_{0}+M^2$, where $M$ denotes the mass (particle number) of a particle in the nonrelativistic boundary field theory. It is assumed to be a positive integer.\footnote{This discreteness originates from the assumption that $\xi$ is a compact coordinate in the Schr\"odinger spacetime $Sch_{D+3}$.} Eq. (\ref{ins9}) can be also casted in the form of the one-dimensional Schr\"odinger equation
\beq \label{ins10}
-\partial_{z}^{2}\psi+\frac{m^2+\frac{(D+2)^2-1}{4}}{z^2}\psi=-\widetilde{q}^2\psi
\eeq
with the help of the substitution $\phi=z^{\frac{D+1}{2}}\psi$. We note, however, that in this case there is no lower bound on the scalar mass $m^2$. The reason is simple: due to the nonrelativistic form of the dispersion relation the condition $\widetilde{q}^2>0$ leads not to the imaginary boundary energy $\omega$, as was valid in the preceding two examples, but only to   $\omega<0$. The nonrelativistic boundary plane-wave excitations remain oscillatory producing no instabilities. No condensate can be formed in the nonrelativistic vacuum, and hence there is no dynamical IR scale generation in the boundary theory. For this reason the $Sch_{D+3}$ spacetime is a reliable background for any scalar mass, and one can use the AdS/CFT correspondence for $m^2<m_{BF}^2=-\frac{(D+2)^2}{4}$. In the next section we draw a similar conclusion by examining the scalar field equation in the global coordinates.
%%%%%%%%%%%%%%%%%%%%%%%%%%%%%%%%%%%%%%%%%%%%%%%%%%%%%%%%%%%%%%%%%%%%%%%%%%%%%%%%%%%%%%%%%%%%%%%%%%%%%%%%%%%
\section{Complex scalar field in global coordinates for $Sch_{D+3}$ spacetime} \label{appendix}
A global coordinate system for $Sch_{D+3}$ spacetime was recently constructed and discussed in \cite{Blau}. Consider a coordinate system $(T,V,R,\vec{X})$ for $Sch_{D+3}$, in terms of which the metric reads
\beq \label{a1}
ds^2=-\frac{dT^2}{R^4}+\frac{1}{R^2}\left(-2dTdV-\omega^2(R^2+\vec{X})dT^2+dR^2+d\vec{X}^2 \right),
\eeq
where $\omega$ is an interpolating frequency parameter and $R\in[0,\infty)$ is the radial coordinate. The metric (\ref{a1}) interpolates smoothly between the Poincar\'e metric ($\omega=0$) and the global metric ($\omega=1$). In what follows we work with a general frequency $\omega$, keeping in mind that the solution for the global coordinates is recovered only after one fixes $\omega=1$.

In this section, we solve following \cite{Blau} the Klein-Gordon equation
\beq \label{a2}
\frac{1}{\sqrt{-g}}\partial_{\mu}\left(\sqrt{-g}g^{\mu\nu}\partial_{\nu}\phi \right)-m_0^2\phi=0
\eeq
for the complex scalar field $\phi$ of mass $m_0$ in the global coordinates. As the coefficients of the metric (\ref{a1}) are independent of the coordinates $T$ and $V$, there are two obvious Killing vector fields $\partial_T$ and $\partial_V$ in the $Sch_{D+3}$ spacetime. Additionally, the metric is symmetric under the rotations in the $\vec{X}$-space. Hence, the scalar eigenmodes with a definite energy $E$, particle number $M$ (assumed to be positive integer) and angular quantum number $L$ can be written as
\beq \label{a3}
\phi_{E,M,L}=e^{-iET}e^{-iMV}Y_L(\Omega_{D-1})\varphi(X)\Phi(R).
\eeq
Here we introduced hyperspherical coordinates in the $\vec{X}$-space, i.e.
\beq \label{a4}
d\vec{X}^2=dX^2+X^2d\Omega_{D-1}^2, \qquad X\in[0,\infty)
\eeq
and $Y_{L}(\Omega_{D-1})$ are spherical harmonics defined on the sphere $S^{D-1}$. The ansatz (\ref{a3}) can now be substituted into the field equation (\ref{a2}), which gives us two separate differential equations for the functions $\varphi(X)$ and $\Phi(R)$
\beq \label{a5}
\partial_{X}^2\varphi+\frac{D-1}{X}\partial_{X}\varphi-M^2\omega^2X^2\varphi-\frac{L(L+D-2)}{X^2}\varphi=-\mathscr{E}\varphi,
\eeq 
\beq \label{a6}
\partial_{R}^2\Phi-\frac{D+1}{R}\partial_{R}\Phi-M^2\omega^2 R^2\Phi-\frac{m^2}{R^2}\Phi=(\mathscr{E}-2ME)\Phi,
\eeq 
where $\mathscr{E}$ is a so far undetermined constant and $m^2=m_0^2+M^2$. The equations (\ref{a5},\ref{a6}) can be rewritten in the form of one-dimensional Schr\"odinger equations by employing the redefinitions $\varphi=X^{\frac{1-D}{2}}\psi$ and $\Phi=R^{\frac{D+1}{2}}\Psi$ yielding the result
\beq \label{a7}
-\partial_{X}^2\psi+M^2\omega^2X^2\psi+\underbrace{(L(L+D-2)+[(D-2)^2-1]/4)}_{-\kappa_X}\frac{\psi}{X^2}=\mathscr{E}\psi,
\eeq 
\beq \label{a8}
-\partial_{R}^2\Psi+M^2\omega^2 R^2\Psi+\underbrace{(m^2+[(D+2)^2-1]/4)}_{-\kappa_{R}}\frac{\Psi}{R^2}=\underbrace{(2ME-\mathscr{E})}_{\bar{E}}\Psi.
\eeq 
Remarkably, both equations define the quantum-mechanical problem of a particle (constrained to a real positive half-line) in a combined inverse square and harmonic potential.

We first consider Eq. (\ref{a7}) with the inverse square potential coupling $\kappa_{X}<0$, which corresponds to a repulsion. This problem was solved by Calogero \cite{Calogero} with the result
\beq \label{a9}
\begin{split}
& \mathscr{E}^{\pm}_{n}=2M\omega (2n\pm a+1), \qquad a=L+\frac{D}{2}-1, \qquad n=0,1,2,..., \\
& \psi_{n}^{\pm}=X^{\pm a+\frac{1}{2}}\exp(-\frac{1}{2}M\omega X^2)L_{n}^{\pm a}(M\omega X^2), \\
\end{split}
\eeq
where $L_{n}^{\pm a}$ denotes a generalized Laguerre polynomial. Due to the harmonic part of the potential the spectrum is discrete and equidistant. Following Calogero, we consider only physically acceptable wave functions, i.e. we require both $|\psi(X)|^2$ and $\psi(X)\psi^{\prime}(X)$ to be continuous. This condition picks out the $\psi^{+}_{n}$ wavefunctions and the corresponding $\mathscr{E}^{+}_n$ branch of the spectrum. The original differential equation (\ref{a5}) thus has the solution
\beq \label{a10}
\varphi^{+}_{n}=X^{\frac{1-D}{2}}\psi_{n}^{+}=X^{L}\exp(-\frac{1}{2}M\omega X^2)L_{n}^{L+\frac{D}{2}-1}(M\omega X^2)
\eeq
in agreement with \cite{Blau}.

\begin{figure}[t]
 \begin{center}
 \includegraphics[height=3.0in]{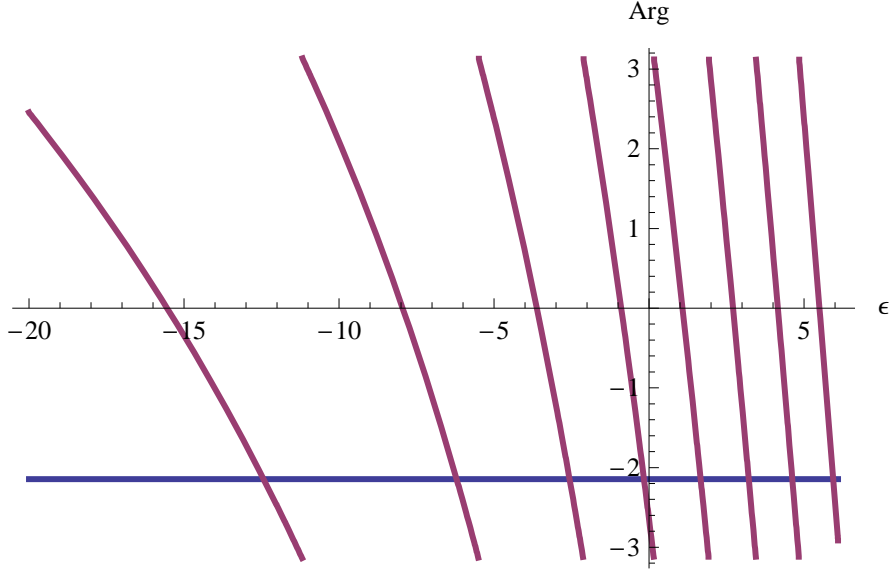}
 \end{center}
 %\vskip -0.7cm 
 \caption{Graphical solution of Eq. (\ref{a13a}) for $u_0=10^{-4}$, $|\nu|=10$ and $M\omega=\frac{1}{4}$. The blue (red) line corresponds to the argument function of the left (right) hand side of Eq. (\ref{a13a}). Points of intersection determine the discrete dimensionless eigenenergies $\epsilon_l=\frac{\bar{E}_l}{4\omega M}$.}
\label{spectrum}
 \end{figure}

Now we turn our attention to the radial equation (\ref{a8}). As expected, up to the harmonic term it is identical with Eq. (\ref{ins10}), derived in Sec. \ref{instability} in the Poincar\'e coordinates. The form of the solution of the differential equation is determined by the value of the inverse square potential coupling $\kappa_{R}$. In particular, for a repulsion and weak attraction $\kappa_R<\frac{1}{4}$, which corresponds to $m^2>m_{BF}^2=-\frac{(D+2)^2}{4}$, the original Calogero solution holds
\beq \label{a11}
\begin{split}
& \bar{E}^{\pm}_{l}=2M\omega (2l\pm a+1), \qquad a=\sqrt{\frac{1}{4}-\kappa_{R}}=\nu, \qquad l=0,1,2,..., \\
& \Psi_{l}^{\pm}=R^{\pm a+\frac{1}{2}}\exp(-\frac{1}{2}M\omega R^2)L_{l}^{\pm a}(M\omega R^2). \\
\end{split}
\eeq
In this case we do not restrict ourselves to the physically acceptable wave functions and consider both branches of Eq. (\ref{a11}). Hence, the radial function $\Phi$ in the original AdS/CFT problem has two branches and reads
\beq \label{a12}
\Phi^{\pm}_{l}=R^{\frac{D+1}{2}}\Psi_{l}^{\pm}=R^{\Delta_{\pm}}\exp(-\frac{1}{2}M\omega R^2)L_{l}^{\pm \nu}(M\omega R^2).
\eeq
The asymptotic behavior of $\Phi^{\pm}_l$ for $R\to 0$ and $R\to\infty$ agrees with findings in the Poincar\'e coordinates. The global energy spectrum is given by
\beq \label{a13}
E^{\pm}_{n,l}=\frac{\mathscr{E}_n+\bar{E}_l}{2M}=\omega\left(2n+2l\pm \nu+L+\frac{D}{2}+1 \right)
\eeq
and has two discrete quantum numbers. On the other hand, for strong attraction $\kappa_R>\frac{1}{4}$ equivalent to $m^2<m_{BF}^2$ the potential in Eq. (\ref{a8}) is truly singular and must be regularized. This case was treated in \cite{Ghosh}, where a cutoff radius $R_0\ll (M\omega)^{-\frac{1}{2}}$ was imposed leading to a boundary condition $\Psi(R_0)=0$ for the wave function $\Psi$. The Hamiltonian of the regularized problem is bounded from below by $\bar{E}_{\text{min}}\approx -\frac{\kappa_R}{R_0^2}$. As shown in \cite{Ghosh}, the energy spectrum $\bar{E}_l$ can be determined from the transcendental equation
\beq \label{a13a}
u_0^{-i|\nu|}=\frac{\Gamma(1-i |\nu|)}{\Gamma(1+i |\nu|)}\frac{\Gamma(\frac{1+i|\nu|}{2}-\frac{\bar{E}}{4\omega M})}{\Gamma(\frac{1-i|\nu|}{2}-\frac{\bar{E}}{4\omega M})}, \qquad \bar{E}>\bar{E}_{\text{min}},
\eeq
where $u_0\equiv M\omega R_0^2$ and $\nu=i|\nu|=\sqrt{\frac{1}{4}-\kappa_R}$. We plot a graphical solution of Eq. (\ref{a13a}) in Fig \ref{spectrum}.

The qualitative features of the energy spectrum can be understood by studying two limits of Eq. (\ref{a13a}). In particular, for energies $\bar{E}_{\text{min}}<\bar{E}\ll 0$ the harmonic term can be neglected in the original Schr\"odinger  equation (\ref{a8}). One is allowed to do so because in this regime the bound state wave functions are well-localized around the origin, where the $1/r^2$ potential dominates over the harmonic potential. Hence, for $\bar{E}_{\text{min}}<\bar{E}\ll 0$ the bound state energies form an almost geometric discrete spectrum, characteristic for the inverse square problem. In fact, by formally expanding Eq. (\ref{a13a}) around $\bar{E}=-\infty$ one obtains the exact geometric scaling of energies $\frac{|\bar{E}_{l+1}|}{|\bar{E}_{l}|}=e^{-\frac{2\pi}{|\nu|}}$.\footnote{We note that this finding is in a disagreement with \cite{Ghosh}, where only one negative energy state in the spectrum was identified.} For $\bar{E}\gg 0$ the bound states wave functions are more sensitive to the large distances where the harmonic potential dominates. In this case the inverse square potential determines only the near-origin behavior of the wave functions. In the limit $\bar{E}\to+\infty$ in Eq. (\ref{a13a}) we get an infinite equidistant spectrum with $\bar{E}_{l+1}-\bar{E}_{l}=4\omega M$. For the moderate values of $\bar{E}$ the spectrum changes its behavior from geometric to equidistant (see Fig. \ref{spectrum}). The arguments presented in this paragraph are close in spirit to \cite{Hammer:2008ra}, where a somewhat similar problem was studied.

%%%%%%%%%%%%%%%%%%%%%%
%%%%%%%%%%%%%%%%%%%%%

Most importantly, in the nonrelativistic AdS/CFT correspondence for $m^2<m_{BF}^2$ the spectrum of global energies
\beq \label{a13b}
E_{n,l}=\frac{\mathscr{E}_n+\bar{E}_l}{2M}=\omega(2n+L+\frac{D}{2})+\frac{\bar{E}_l}{2M}
\eeq
 remains real. For this reason we conclude that  there appears no instability in the solution as one crosses the BF bound. This is in a stark contrast with the solution of the scalar field equation in the global coordinates for $AdS_{d+1}$ \cite{BF}. In this case the global energy is actually conserved only if its flux at infinity vanishes. This leads to the quantization condition
\beq \label{a100}
E_n=2n+\Delta_{\pm}+L, \qquad n=0,1,2,...
\eeq
According to our findings in the Poincar\'e coordinates, the scaling dimension $\Delta_{\pm}=\frac{d}{2}\pm\sqrt{\frac{d^2}{4}+m^2}$ becomes complex below the relativistic BF bound and the global energy $E_n$ acquires an imaginary part signaling a presence of instability.

%%%%%%%%%%%%%%%%%%%%%%%%%%%%%%%%%%%%%%%%%%%%%%%%%%%%%%%%%%%%%%%%%%%%%%%%%%%%%%%%%%%%%%%%%%%%%%%%%%%%%%%%%%%
\section{Inverse square potential and holographic RG flows} \label{flows}
In the context of the relativistic AdS/CFT correspondence it was shown in \cite{Klebanov} that, if the mass of a (complex) scalar bulk field $\phi$ lies in the interval
\beq \label{dg1}
-\frac{d^2}{4}<m^2<-\frac{d^2}{4}+1,
\eeq
then a single gravity theory in the bulk describes two different conformal field theories on the boundary which we call $\text{CFT}_+$ and $\text{CFT}_-$. The CFT operator $O$ dual to the scalar field $\phi$ has the scaling dimension
\beq \label{dg2}
\Delta_{\pm}=\frac{d}{2}\pm\sqrt{\frac{d^2}{4}+m^2}
\eeq
in these two conformal field theories. The near-boundary asymptotic of the field $\phi$ reads
\beq \label{dg3}
\phi(z)=c_- z^{\Delta_-}+c_+z^{\Delta_+},
\eeq
where $c_-$ and $c_+$ are proportional to the source $J$ coupled to $O$ and the condensate $\langle O \rangle$ in $\text{CFT}_+$ perturbed by $\int d^d x (J^{\dagger}O+ J O^{\dagger})$. In the perturbed $\text{CFT}_-$ the roles of the coefficients $c_-$ and $c_+$ are interchanged: $c_-$ is identified with the condensate, while $c_+$ with the source \cite{Klebanov}.

In this section we demonstrate how the two CFTs arise in the equivalent one-dimensional problem of quantum mechanics with an inverse square potential (a similar construction was made recently in \cite{Kaplan}) and construct RG flows between $\text{CFT}_+$ to $\text{CFT}_-$. Additionally, we discuss the case of the nonrelativistic version of AdS/CFT, where one can go below the BF bound without experiencing any instability. In particular, we construct RG flows and compute the two-point scalar correlation function for $m^2<m_{BF}^2$ in the nonrelativistic holography.

\subsection{Relativistic holography}
We consider the regime of $m^2>m_{BF}^2$ and $q^2>0$, where the solution of Eq. (\ref{ins4}) regular in the bulk is given by
\beq \label{dg4}
\phi_{q}(z)=\frac{z^{d/2}K_{\nu}(qz)}{z_0^{d/2}K_{\nu}(qz_0)}, \qquad \nu=\sqrt{\frac{d^2}{4}+m^2}.
\eeq
Here we introduced the IR bulk cutoff $z_0$, which corresponds to the UV momentum cutoff in the boundary field theory, and normalized the solution so that $\phi_q(z_0)=1$.  Albeit the first condition for instability $q^2>0$ from Sec. \ref{instability} is fulfilled by the solution (\ref{dg4}), the second condition due to Calogero is not satisfied by it. For $m^2>m_{BF}^2$ the associated $\psi_q=z^{(1-d)/2}\phi_q$ is not a physically acceptable bound state wave function in the inverse square potential problem. Hence, one is in the stable regime.

In the interval (\ref{dg1}) the single bulk solution (\ref{dg4}) describes two different conformal boundary field theories. The RG flow from $\text{CFT}_+$ to $\text{CFT}_-$ can be obtained by turning $ O^{\dagger}O$ interaction in the boundary field theory \cite{Gubser} (for the AdS/CFT treatment see \cite{Witten}). Some insight into this subtlety can be gained by considering the equivalent Schr\"odinger  equation (\ref{ins5}) in the regime $y\equiv zq\ll 1$ and $\kappa<\kappa_{cr}$. In this domain the general complex solution of Eq. (\ref{ins5}) reads
\beq \label{dg5}
\psi(z)=c_-z^{1/2-\nu}+c_+ z^{1/2+\nu}=\widetilde{c}_-y^{1/2-\nu}+\widetilde{c}_+ y^{1/2+\nu},\qquad \nu=\sqrt{\frac{1}{4}-\kappa}
\eeq
with $c_-, c_+\in\mathbb{C}$ and $\widetilde{c}_{\pm}\equiv\frac{c_{\pm}}{q^{1/2\pm\nu}}$. Notably, only in the interval (\ref{dg1}) the general solution (\ref{dg5}) is square integrable around the origin. Thus, in the quantum mechanical language it is the property of square integrability around the origin which determines whether the solution (\ref{dg4}) is normalizable or not. This point of view complements the original arguments of Breitenlohner and Freedman \cite{BF} based on the finiteness of energy in the $AdS$ spacetime or the argument of Klebanov and Witten \cite{Klebanov} based on the finiteness of the action in the Euclidean $AdS$ space. 

 The inverse square potential is singular at $z=0$ and in its own gives an ill-defined quantum mechanical problem. It must be regularized at the origin and this can be achieved in various ways \cite{Beane:2000wh,Barford:2002je,Bawin:2003dm,Camblong,Ho,Braaten,Barford:2004fz,Hammer:2005sa,MS}. We choose to regularize the problem by extending it to the full real $z$-line, cutting the potential off for $|z|<z_0$ and introducing a localized $\delta$-function potential (counterterm) at the origin. As will be illustrated in the rest of this subsection, this procedure corresponds to the inclusion of the double-trace $O^{\dagger}O$ boundary contact term in the AdS/CFT correspondence \cite{Witten}.

\begin{figure}
\includegraphics[height=3in]{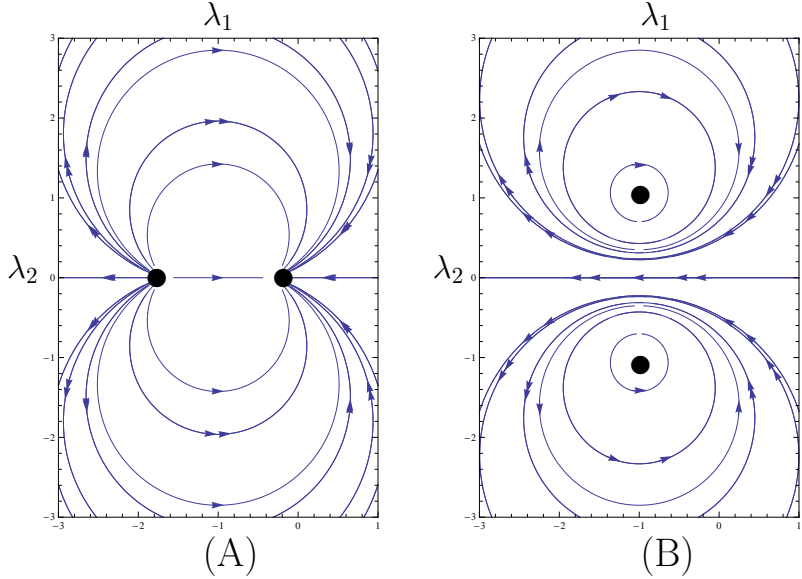} 
\caption{The phase portraits of  the RG flows of the contact complex coupling $\lambda=\lambda_{1}+i\lambda_{2}$ in the one-dimensional inverse square potential problem: (A) undercritical regime $\kappa=\frac{1}{8}<\kappa_{cr}$, (B) overcritical regime $\kappa=\frac{1}{2}>\kappa_{cr}$. Arrows denote the direction towards the UV. In AdS/CFT the case (A) corresponds to $m^2>m_{BF}^2$, while (B) appears for $m^2<m_{BF}^2$.}
\label{bos}
\end{figure}

The modified regularized potential reads
\beq \label{dg6}
V(z)=\left\{ \begin{array}{c} -\frac{\lambda}{z_0}\delta(z), \qquad |z|<z_0,   \\ -\frac{\kappa}{z^2}, \qquad |z|>z_0, \\  \end{array} \right.
\eeq
where the coordinate $z\in\mathbb{R}$ in the regularized quantum mechanical one-dimensional problem, and $z_0$ serves as an IR position cutoff. $\lambda$ is a dimensionless coupling constant. Here we find a general solution of the regularized problem with the negative energy $E=-q^2<0$. We concentrate our attention on the domain $z>0$, $y\equiv z q\ll 1$, where the quantum mechanical wave function is given by
\beq \label{dg7}
\psi(y)=\left\{ \begin{array}{c} e^{-y}+D e^{y}, \qquad 0<y<y_0\equiv q z_0,   \\ \mathscr{N}(\widetilde{c}_-y^{1/2-\nu}+\widetilde{c}_+y^{1/2+\nu}), \qquad y>y_0. \\  \end{array} \right.
\eeq
Note that $\psi(y)$ must be an even function because an odd $\psi(y)$, which is also allowed by the even potential (\ref{dg6}), yields necessarily a vanishing counterterm $\lambda$. Now the coefficient $D$ can be easily expressed from the continuity of the wave function $\psi(y)$ and the known discontinuity of its first derivative at $y=0$. Specifically, $D$ is related to the coupling constant $\lambda$ by
\beq \label{dg7a}
\lambda=2y\frac{1-D}{1+D}.
\eeq
The normalization constant $\mathscr{N}$ can be determined from the continuity condition $\psi(y)|_{y\to y_0-0}=\psi(y)|_{y\to y_0+0}$
\beq \label{dg8}
e^{-y_0}+D e^{y_0}\approx 1+D=\mathscr{N}(\widetilde{c}_- y_0^{1/2-\nu}+\widetilde{c}_+ y_0^{1/2+\nu}).
\eeq
Additionally, by matching the derivatives $\frac{d}{dy}\psi|_{y\to y_0-0}=\frac{d}{dy}\psi|_{y\to y_0+0}$ we obtain
\beq \label{dg8a}
-e^{-y_0}+D e^{y_0}\approx -1+D=\mathscr{N}\left[\widetilde{c}_-(1/2-\nu) y_0^{-1/2-\nu}+\widetilde{c}_+(1/2+\nu) y_0^{-1/2+\nu}\right].
\eeq
Finally, substituting Eqs. (\ref{dg8}) and (\ref{dg8a}) into Eq. (\ref{dg7a}) one arrives at
\beq \label{dg9}
\lambda(t)=-1+2\nu\frac{e^{\nu t}-C e^{-\nu t}}{e^{\nu t}+C e^{-\nu t}}, \qquad C\equiv \frac{\widetilde{c}_+}{\widetilde{c}_-},
\eeq
where we introduced $t\equiv -\ln(q z_0)$. This expression can be interpreted as the RG flow of the contact coupling $\lambda$ as a function of the logarithmic RG scale $t$. The dimensionless parameter $C$ which we allow to be complex is considered to be a fixed constant during the RG evolution. It determines in general complex initial condition $\lambda(t=0)$ via the M\"obius (linear fractional) transformation. The RG flow (\ref{dg9}) solves the inhomogeneous Riccati differential equation
\beq \label{dg10}
\partial_t \lambda=-\frac{\lambda^2}{2}-\lambda-2\kappa=-\frac{1}{2}\left(\lambda+1-2\nu \right)\left(\lambda+1+2\nu \right)
\eeq
and possesses the UV fixed point $\lambda_{UV}=-1+2\nu$ (corresponds to the boundary $\text{CFT}_-$ in AdS/CFT) and the IR fixed point $\lambda_{IR}=-1-2\nu$ (corresponds to the $\text{CFT}_+$ in AdS/CFT). The phase portrait of the RG evolution in the complex $\lambda$ plane can be found in Fig. \ref{bos}(A).
%%%%%%%%%%%%%%%%%%%%%%%%%%%%%%%
%%%%%%%%%%%%%%%%%%%%%%%%%%%%%%%
The RG trajectories are arcs of circles of radius $R=2\nu |\frac{C}{\text{Im}C}|$. 

The extension to a complex coupling $\lambda$ provides a deeper mathematical understanding of renormalization in the nonrelativistic quantum mechanical problem with an inverse square potential. It was introduced and motivated in \cite{MS}. Physically, the generalization to the complex $\lambda$ opens an inelastic channel in the quantum mechanical scattering and converts bound states to metastable resonances. Here we note that one can further investigate this extension using holography by allowing bulk solutions with generally complex $c_{-}$ and $c_{+}$ in the asymptotic expression (\ref{dg3}). Complex generalization also plays an important role in the next subsection, where the nonrelativistic case is discussed.

Of special physical interest, however,  is the real domain of the coupling $\lambda$, which corresponds to $C\in\mathbb{R}$ in Eq. (\ref{dg9}). Notably, both fixed points $\lambda_{UV}$ and $\lambda_{IR}$ lie on the real axis. If one tunes the initial condition $\lambda(t=0)$ to the real line, then the RG flows remain on the real line. In the AdS/CFT correspondence, information about both conformal field theories is contained in the solution (\ref{dg4}). The detailed procedure of how the values of the source $J$ coupled to the operator $O$ and the condensate $\langle O \rangle$ are extracted from the asymptotic form (\ref{dg3}) can be found in \cite{Klebanov, Kaplan}. For sake of comparison with the nonrelativistic result which we derive in the next subsection, note that the two-point function $\langle O O^{\dagger}\rangle$ in momentum space is proportional to $q^{2\nu}$ in the boundary $\text{CFT}_{+}$ and $q^{-2\nu}$ in the boundary $\text{CFT}_{-}$.
%%%%%%%%%%%%%%%%%%%%%%%%%%%%%%%%%%%%%%%%%%%%%%%%%%%%%%%%%%%%%%%%%%%%%%%%%%%%%%%
%%%%%%%%%%%%%%%%%%%%%%%%%%%%%%%%%%%%%%%%%%%%%%%%%%%%%%%%%%%%%%%%%%%%%%%%%%%%%%%
\subsection{Nonrelativistic holography}
The ideas presented in the last subsection can be straightforwardly applied to the case of the nonrelativistic AdS/CFT correspondence. All the derived results still hold provided we substitute $d\to D+2$ and $q\to\widetilde{q}$. Moreover, as was pointed out in Sec. \ref{instability}, we can go below the nonrelativistic (spurious) BF bound $m_{BF}^2=-\frac{(D+2)^2}{4}$ without any instability in the $Sch_{D+3}$ spacetime.

In this subsection we concentrate our attention to the interesting regime $m^2<m_{BF}^2$ for $\widetilde{q}^2>0$, where the regular bulk solution of Eq. (\ref{ins9}) is given by
\beq \label{dg11}
\phi_{\widetilde{q}}(z)=\frac{z^{(D+2)/2}K_{i|\nu|}(\widetilde{q}z)}{z_0^{(D+2)/2}K_{i|\nu|}(\widetilde{q}z_0)}, \qquad \nu=\sqrt{\frac{(D+2)^2}{4}+m^2}.
\eeq
This solution is real and normalized as $\phi_{\widetilde{q}}(z_0)=1$.

In order to gain some intuition we first solve the equivalent regularized one-dimensional Schr\"odinger equation (\ref{ins10}) in the regime $\kappa>\kappa_{\text{cr}}=\frac{1}{4}$. For $z>0$, $y\equiv \widetilde{q}z\ll 1$ the general solution reads
\beq \label{dg12}
\psi(z)=c_- z^{1/2-i|\nu|}+c_+ z^{1/2+i |\nu|}=\widetilde{c}_- y^{1/2-i|\nu|}+\widetilde{c}_+ y^{1/2+i |\nu|}, \qquad \nu=\sqrt{\frac{1}{4}-\kappa},
\eeq
where $\widetilde{c}_{\pm}\equiv\frac{c_{\pm}}{\widetilde{q}^{1/2\pm \nu}}$.
Here $\psi(z)$ is square integrable around the origin for any $c_-, c_+\in \mathbb{C}$, hence the general solution (\ref{dg11}) is normalizable. Following the same steps as in the last subsection, we construct the RG trajectories of the contact emergent coupling $\lambda$ which take the form
\beq \label{dg13}
\lambda(t)=-1+2i|\nu|\frac{e^{i|\nu| t}-C e^{-i|\nu| t}}{e^{i|\nu| t}+C e^{-i|\nu| t}}, \qquad C=\frac{\widetilde{c}_+}{\widetilde{c}_-}.
\eeq
The RG flow solves Eq. (\ref{dg10}) and has two complex fixed points $\lambda_{\pm}=-1\pm 2i |\nu|$. Its phase portrait is depicted in Fig. \ref{bos}(B). In the complex plane the RG trajectories form closed circles of radius $R=4|\frac{C \nu}{1-|C|^2}|$. The real $\lambda$-line is a separatrix of the two complex fixed points, and for the real initial condition\footnote{which leads to $C=e^{i\varphi}$ in Eq. (\ref{dg13}), i.e. $C$ must be a pure complex phase.} $\lambda(t=0)\in\mathbb{R}$  the RG flow remains on the real line. In this regime the renormalization of the coupling $\lambda$ exhibits an infinite (unbounded) limit cycle\footnote{Strictly speaking, to find a finite (bounded) limit cycle, i.e. a closed curve in the space of running couplings, one needs at least two real coupling constants connected by the RG flow equations. In the case of a single coupling constant, an infinite (unbounded) limit cycle appears only if there are periodic discontinuities in the RG evolution, and the RG `` flows through'' infinity. An elegant geometric description of the infinite limit cycle can be provided on the Riemann sphere \cite{MS}.} and periodically traverses the real $\lambda$-line. The continuous scale symmetry of the classical inverse square potential is broken to the discrete subgroup $Z$ by a quantum anomaly.

In the rest of this subsection we demonstrate that for $m^2<m_{BF}^2$ the real gravity solution (\ref{dg11}) is dual to a nonrelativistic boundary field theory  with an unbounded limit cycle. To this end, first, we find from the near-boundary asymptotic form (\ref{dg3}) of the solution (\ref{dg11}) that
\beq \label{dg14}
C=\frac{\widetilde{c}_+}{\widetilde{c}_-}=\frac{\Gamma(-i|\nu|)}{\Gamma(i|\nu|)}\left(\frac{1}{2} \right)^{2i|\nu|},
\eeq
which is a pure complex phase. This gives rise to the real initial condition $\lambda(t=0)$ in the equivalent inverse square potential problem and tunes the RG flow to the real limit cycle. Second, using the standard AdS/CFT machinery, we calculate the two-point function $\langle O O^{\dagger} \rangle$ of the operator $O$ dual to the scalar field $\phi$ with $m^2<m_{BF}^2$ in the nonrelativistic holography. The two-point correlator can be extracted from the quadratic part of the on-shell action $S_{\text{on-shell}}$, which can be written as the boundary integral
\beq \label{dg15}
S_{\text{on-shell}}[\phi_0, \phi^{*}_0]=-\int dX \sqrt{-g} g^{zz}\phi^{*}(X,z)\partial_z \phi(X,z)|_{z=z_0}
\eeq
with $X=\{t,\xi,\vec{x}\}$. The general on-shell field $\phi(X,z)$ can be now decomposed into the Fourier modes
\beq \label{dg16}
\phi(X,z)=\int d^{D+2}Q \phi_0(Q,z_0) \phi_{\widetilde{q}}(z)e^{iQ\cdot X}, \qquad Q=\{\omega, M, \vec{q} \}.
\eeq
Using this representation, the on-shell action can be conveniently rewritten as
\beq \label{dg17}
S_{\text{on-shell}}[\phi_0, , \phi^{*}_0]=-\int d^{D+2}Q \phi_0^{*}(Q,z_0)\mathscr{F}(\widetilde{q},z_0)\phi_0(Q,z_0)
\eeq
with the flux factor
\beq \label{dg18}
\mathscr{F}(\widetilde{q},z_0)=\lim_{z\to z_0}\sqrt{-g}g^{zz}\phi^{*}_{\widetilde{q}}(z)\partial_z \phi_{\widetilde{q}}(z).
\eeq
The two-point function can now easily be expressed in terms of the flux factor
\beq \label{dg19}
\langle O(Q_1) O^{\dagger}(Q_2) \rangle=-\frac{\delta}{\delta \phi_0^*(Q_1)}\frac{\delta}{\delta \phi_0(Q_2)}S_{\text{on-shell}}[\phi_0, , \phi^{*}_0]=(2\pi)^{D+2}\delta(Q_1-Q_2)\mathscr{F}(\widetilde{q},z_0).
\eeq
Substituting Eq. (\ref{dg11}) into Eq. (\ref{dg18}) we obtain
\beq \label{dg20}
\mathscr{F}(\widetilde{q},z_0)=z_0^{-D-1}\partial_z \frac{z^{(D+2)/2}K_{i|\nu|}(\widetilde{q}z)}{z_0^{(D+2)/2}K_{i|\nu|}(\widetilde{q}z_0)} |_{z=z_0}.
\eeq
This can be evaluated by introducing the near-boundary asymptotic of the Bessel function
\beq \label{dg21}
K_{i|\nu|}(\widetilde{q}z)=a_- (\widetilde{q}z)^{-i|\nu|}+a_+ (\widetilde{q}z)^{i|\nu|}, \quad a_+=a_-^*=2^{-1-i|\nu|}\Gamma(-i|\nu|)\equiv |a_+|e^{i\alpha}
\eeq
into Eq. (\ref{dg20})
\beq \label{dg22}
\begin{split}
\mathscr{F}(\widetilde{q},z_0)=&z_0^{-D-1}\partial_z \frac{a_-(\widetilde{q}z)^{-i|\nu|+\frac{D+2}{2}}+a_+(\widetilde{q}z)^{i|\nu|+\frac{D+2}{2}}}{a_-(\widetilde{q}z_0)^{-i|\nu|+\frac{D+2}{2}}+a_+(\widetilde{q}z_0)^{i|\nu|+\frac{D+2}{2}}} |_{z=z_0} \\
&=z_{0}^{-D-2}(D/2+1)-z_{0}^{-D-2}|\nu|\tan \left\{|\nu|\ln (\widetilde{q}z_0)+\alpha\right\}.
\end{split}
\eeq
The first term is a contact contribution and can be subtracted by a proper boundary counterterm. Thus, the two-point function in momentum space is given by
\beq \label{dg23}
\langle O O^{\dagger} \rangle\sim\tan \left\{|\nu|\ln\widetilde{q}+\underbrace{|\nu|\ln z_0+\alpha}_{\gamma(z_0)}\right\}
\eeq
and has the following properties:
\begin{itemize}
\item $\langle O O^{\dagger} \rangle$ is log-periodic in $\widetilde{q}$ with the period $T=\frac{\pi}{|\nu|}$.
\item The infinite series of simple pole divergences of the two-point function indicates that the boundary field $O$ describes infinitely many stable particles with energies
\beq \label{dg24}
\omega_n=-\frac{1}{2M}\exp\left(-\frac{2 \pi n}{|\nu|}+\frac{\pi-2\gamma(z_0)}{|\nu|} \right), \quad n\in\mathbb{Z}.
\eeq
The spectrum is infinite with the accumulation point at $\omega=0$ as $n\to\infty$. It exhibits the geometric behavior
\beq \label{dg25}
\frac{\omega_{n+1}}{\omega_n}=\exp\left(-\frac{2\pi}{|\nu|} \right).
\eeq
From the form of the energy spectrum one can infer, that the continuous scale symmetry is broken to the discrete subgroup $Z$, and we are dealing with a limit cycle solution of the renormalization group.
\item As every limit cycle solution has to be defined with a physical UV momentum cutoff \cite{BH}, the two-point function (\ref{dg23}) explicitly depends on $z_0$ through the angle $\gamma(z_0)$. In the RG language $\gamma(z_0)$ determines the ultraviolet initial condition on the limit cycle trajectory.
\end{itemize}

There is a well-known subtlety in the calculation of the two-point function in the relativistic AdS/CFT correspondence. Only by using the normalized solution (\ref{dg4}) and expanding both the numerator and the denominator of the relativistic version of the flux factor formula (\ref{dg20}) one obtains the correct normalized two-point function, which is consistent with the Ward identity \cite{Freedman98}. We stress that in the nonrelativistic calculation presented above it is absolutely crucial to use the normalized bulk solution (\ref{dg11}) and follow the correct prescription. Without proper normalization one would obtain $\langle O O^{\dagger} \rangle\sim\sin \left\{|\nu|\ln(\widetilde{q}z_0)+\alpha\right\}$.

The limit cycle solution appears in different nonrelativstic quantum mechanical problems \cite{BH}. One prominent example is the Efimov effect \cite{Efimov} for three identical bosons interacting through a pointlike potential tuned to the unitarity point.\footnote{At the unitarity point the quantum two-body problem has a zero-energy shallow bound state and a scattering cross section saturates the unitarity bound \cite{Bloch, giorgini}.} Remarkably, an infinite geometric three-body spectrum is developed in this system signaling the limit cycle RG behavior, i.e. the nonrelativistic quantum scale anomaly \cite{Camblong}. The RG period of this limit cycle is $T=\frac{\pi}{s_0}$, where the so-called Efimov parameter is $s_0\approx 1.0062$. The first experimental signatures of the Efimov effect were recently observed in experiments with cold bosonic atoms \cite{Kraemer}.  We speculate that the Efimov effect can be studied with the nonrelativistic AdS/CFT correspondence by incorporating bulk scalars with $m^2<m_{BF}^2$. In particular, the local atom-dimer composite scalar operator $O=\psi\phi$ has the complex scaling dimension
\beq \label{dg26}
\Delta_{\pm}=\frac{5}{2}\pm is_{0}
\eeq
in the three-dimensional Efimov problem \cite{Nishida1}. In the light of our proposal, this operator should be dual to the bulk scalar with $\nu=\pm i s_{0}$, i.e. with $m^2<m_{BF}^{2}$. It would be interesting to calculate different n-point correlation functions\footnote{The prescription for calculation of the n-point correlators in the nonrelativistic AdS/CFT correspondence was  given recently in \cite{Fuertes, Volovich}.} involving this bulk scalar in $Sch_{D+3}$ for $D=3$ and to compare the result with the known field-theoretical calculations of the scattering amplitudes in the Efimov physics \cite{BH}.

Another, more simple example, where limit cycles appear, is quantum mechanics in general $D$ spatial dimensions with an inverse square potential \cite{Case,Furlan,Kolomeisky,Gupta,Beane:2000wh,Barford:2002je,Bawin:2003dm,Ho,Braaten,Barford:2004fz,Hammer:2005sa,Hammer:2008ra, MS}. In the nonconformal phase, i.e. for $\kappa>\kappa_{cr}=\frac{(D-2)^2}{4}$, the composite operator $O=\psi\psi$ acquires a complex scaling dimension. If a gravity dual of this problem can be constructed, the field $O$ should be dual to a bulk scalar with $m^2<m_{BF}^2$.

We are not aware of the field-theoretical calculations of the two-point function of the composite operators introduced in the preceding two paragraphs. However, it is reassuring that the renormalization group studies \cite{BH, MS} of the relevant couplings in both quantum theories reveal periodic dependence on the logarithmic RG scale $t$ of the form\footnote{Specifically, $|\nu|=s_{0}$ for the Efimov effect and $|\nu|=\sqrt{\kappa-\frac{(D-2)^2}{4}}$ in the inverse square potential problem.} $\sim\tan(|\nu| t)$, which is consistent with our result (\ref{dg23}).

%%%%%%%%%%%%%%%%%%%%%%%%%%%%%%%%%%%%%%%%%%%%%%%%%%%%%%%%%%%%%%%%%%%%
\section{Conclusion and outlook}
In this work we revisited the problem of a free complex scalar field in the Poincar\'e coordinates for the anti-de Sitter and Schr\"odinger background spacetimes by exploiting its mathematical equivalence to the well-understood problem of quantum mechanics with an inverse square potential in one spatial dimension. With the help of this equivalence it was demonstrated that, due to the nonrelativistic form of the  boundary dispersion relation, there is no need for the mass stability bound in the nonrelativistic AdS/CFT correspondence. We arrived to the same conclusion by solving the problem in the global coordinates for $Sch_{D+3}$.

In the domain where a single bulk theory describes two different conformal field theories we related the RG flows between the two CFTs to the RG evolution of the emergent contact coupling constant in the inverse square potential problem. We argued that for a deeper mathematical understanding the RG flows can be extended to the complex values of the coupling constant and we motivated this generalization.

Finally, the scalar two-point correlator for $m^2<m_{BF}^2$ was computed in the nonrelativistic holography. Most importantly, the two-point function turned out to depend explicitly on the momentum cut-off, thus violating nonrelativistic continuous scale symmetry. It was demonstrated, however, that it is symmetric under the discrete scale symmetry subgroup. For this reason we concluded that for $m^2<m^2_{BF}$ the nonrelativistic holography describes a quantum field theory with a quantum mechanical scale anomaly, manifested by the RG limit cycle scaling. As the well-known quantum-mechanical Efimov effect for three equivalent bosons provides a paradigmatic realization of limit cycles in atomic and nuclear physics, we propose that cold bosons at unitarity and the Efimov effect in particular can be studied in the framework of the nonrelativistic AdS/CFT correspondence.

In this paper we argued that there is no mass stability bound for the free scalar field in the nonrelativistic AdS/CFT correspondence. It is natural to ask the question whether our finding still holds even for the interacting scalar theory in the bulk. As an example, let us modify the free action (\ref{ins8}) in $Sch_{D+3}$ spacetime by adding the interaction part
\beq \label{inter1}
S_{\text{int}}[\phi, \phi^{*}]=-\frac{\alpha}{2}\int dz dt d\xi d^{D}x\sqrt{-g}(\phi^{*}\phi)^2.
\eeq
As in Sec. \ref{instability}, the scalar field equation can still be mapped onto the one-dimensional quantum mechanical ''Schr\"odinger`` equation
\beq \label{inter2}
-\partial_{z}^{2}\psi(Q,z)+\frac{m^2+\frac{(D+2)^2-1}{4}}{z^2}\psi(Q,z)+\alpha z^{D-1}\psi^{*}\psi^2(X,z)=-\widetilde{q}^2\psi(Q,z),
\eeq
where $X\equiv\{t,\xi,\vec{x}\}$ and $Q\equiv\{\omega,M,\vec{q}\}$ and the interaction term is expressed in the position space, where it is local. The equation is nonlinear, and the analytic solution is difficult to obtain. Physically, we expect the nonlinear term to modify the spectrum, but not to change its reality property.  If the energy spectrum of Eq. (\ref{inter2}) is real, the argument presented in Sec. \ref{instability} still holds, and there is no mass stability bound even in the presence of interactions. The proper inclusion of interactions can be done perturbatively using diagrammatic techniques of AdS/CFT, and we defer this problem to future.

\emph{Acknowledgments} -- It is our pleasure to acknowledge discussions with D.~Son which inspired this work. We are thankful to J.~Barb\'on, M.~Blau, C.~Fuertes, K.~Gupta, H.~Hammer and M.~Rangamani for communication. We are grateful to A.~Hebecker and T.~Proch\'azka for valuable critical remarks to the manuscript. The author is supported by KTF.
%%%%%%%%%%%%%%%%%%%%%%%%%%%%%%%%%%%%%%%%%%%%%%%%%%%%%%%%%%%%%%%%%%%%%%
%\appendix


\begin{thebibliography}{99}
\bibitem{AdSCFT}
  J.~M.~Maldacena,
  %``The large N limit of superconformal field theories and supergravity,''
  Adv.\ Theor.\ Math.\ Phys.\  {\bf 2} (1998) 231
  [Int.\ J.\ Theor.\ Phys.\  {\bf 38} (1999) 1113];
  %[arXiv:hep-th/9711200].
  %%CITATION = IJTPB,38,1113;%%
  S.~S.~Gubser, I.~R.~Klebanov and A.~M.~Polyakov,
  %``Gauge theory correlators from non-critical string theory,''
  Phys.\ Lett.\  B {\bf 428} (1998) 105;
  %[arXiv:hep-th/9802109].
  %%CITATION = PHLTA,B428,105;%%
  E.~Witten,
 % ``Anti-de Sitter space and holography,''
  Adv.\ Theor.\ Math.\ Phys.\  {\bf 2} (1998) 253.
  %[arXiv:hep-th/9802150].
  %%CITATION = 00203,2,253;%%
 \bibitem{Son}
  D.~T.~Son,
  %``Toward an AdS/cold atoms correspondence: a geometric realization of the
  %Schroedinger symmetry,''
  Phys.\ Rev.\  D {\bf 78}, 046003 (2008).
  %[arXiv:0804.3972 [hep-th]].
  %%CITATION = PHRVA,D78,046003;%%
\bibitem{McGreevy}
  K.~Balasubramanian and J.~McGreevy,
  %``Gravity duals for non-relativistic CFTs,''
  Phys.\ Rev.\ Lett.\  {\bf 101}, 061601 (2008).
  %[arXiv:0804.4053 [hep-th]].
  %%CITATION = PRLTA,101,061601;%%

\bibitem{Kachru}
  S.~Kachru, X.~Liu and M.~Mulligan,
  %``Gravity Duals of Lifshitz-like Fixed Points,''
  Phys.\ Rev.\  D {\bf 78} (2008) 106005.
  %[arXiv:0808.1725 [hep-th]].
  %%CITATION = PHRVA,D78,106005;%%
\bibitem{Bloch}
  I.~Bloch, J.~Dalibard and W.~Zwerger,
  %``Many-Body Physics with Ultracold Gases,''
  Rev.\ Mod.\ Phys.\ {\bf 80}, 885 (2008).
  %[arXiv:0704.3011 [cond-mat.other]]
\bibitem{giorgini}
  S.~Giorgini, L.~P.~Pitaevskii and S.~Stringari,
  %``Theory of ultracold atomic Fermi gases,''
  Rev.\ Mod.\ Phys.\  {\bf 80}, 1215 (2008).
\bibitem{Nishida}
  Y.~Nishida and D.~T.~Son,
  %``Nonrelativistic conformal field theories,''
  Phys.\ Rev.\  D {\bf 76}, 086004 (2007).
  %%CITATION = PHRVA,D76,086004;%%

\bibitem{Hartnoll}
  S.~A.~Hartnoll,
  %``Lectures on holographic methods for condensed matter physics,''
  Class.\ Quant.\ Grav.\  {\bf 26}, 224002 (2009).
  %[arXiv:0903.3246 [hep-th]].
  %%CITATION = CQGRD,26,224002;%%
\bibitem{FB}
  J.~L.~F.~Barbon and C.~A.~Fuertes,
  %``On the spectrum of nonrelativistic AdS/CFT,''
  JHEP {\bf 0809}, 030 (2008).
  %[arXiv:0806.3244 [hep-th]].
  %%CITATION = JHEPA,0809,030;%%
\bibitem{Kaplan}
  D.~B.~Kaplan, J.~W.~Lee, D.~T.~Son and M.~A.~Stephanov,
  %``Conformality Lost,''
  arXiv:0905.4752 [hep-th].
  %%CITATION = ARXIV:0905.4752;%%
\bibitem{Dobrev}
  V.~K.~Dobrev,
  %``Intertwining operator realization of the AdS/CFT correspondence,''
  Nucl.\ Phys.\  B {\bf 553}, 559 (1999).
  %[arXiv:hep-th/9812194].
  %%CITATION = NUPHA,B553,559;%%
\bibitem{Klebanov}
  I.~R.~Klebanov and E.~Witten,
  %``AdS/CFT correspondence and symmetry breaking,''
  Nucl.\ Phys.\  B {\bf 556}, 89 (1999).
  %[arXiv:hep-th/9905104].
  %%CITATION = NUPHA,B556,89;%%
\bibitem{BF}
  P.~Breitenlohner and D.~Z.~Freedman,
  %``Stability In Gauged Extended Supergravity,''
  Annals Phys.\  {\bf 144}, 249 (1982);
  %%CITATION = APNYA,144,249;%%
  P.~Breitenlohner and D.~Z.~Freedman,
  %``Positive Energy In Anti-De Sitter Backgrounds And Gauged Extended
  %Supergravity,''
  Phys.\ Lett.\  B {\bf 115}, 197 (1982).
  %%CITATION = PHLTA,B115,197;%%
 \bibitem{Case}
  K.~M.~Case,
  %``Singular potentials,''
  Phys.\ Rev.\  {\bf 80}, 797 (1950).
  %%CITATION = PHRVA,80,797;%%
\bibitem{Furlan}
  V.~de Alfaro, S.~Fubini and G.~Furlan,
  %``Conformal Invariance In Quantum Mechanics,''
  Nuovo Cim.\  A {\bf 34}, 569 (1976).
  %%CITATION = NUCIA,A34,569;%%
  \bibitem{Kolomeisky}
  E.~Kolomeisky and J.~P.~Straley,
  Phys.\ Rev.\  B {\bf 46}, 12664 (1992).
\bibitem{Gupta}
  K.~S.~Gupta and S.~G.~Rajeev,
  %``Renormalization in quantum mechanics,''
  Phys.\ Rev.\  D {\bf 48}, 5940 (1993).
  %[arXiv:hep-th/9305052].
 \bibitem{Beane:2000wh}
  S.~R.~Beane, P.~F.~Bedaque, L.~Childress, A.~Kryjevski, J.~McGuire and U.~v.~Kolck,
  %``Singular Potentials and Limit Cycles,''
  Phys.\ Rev.\  A {\bf 64}, 042103 (2001).
  %[arXiv:quant-ph/0010073].
  %%CITATION = PHRVA,A64,042103;%%
 \bibitem{Barford:2002je}
  T.~Barford and M.~C.~Birse,
  %``A renormalisation group approach to two-body scattering in the presence  of
  %long-range forces,''
  Phys.\ Rev.\  C {\bf 67}, 064006 (2003).
  %[arXiv:hep-ph/0206146].
  %%CITATION = PHRVA,C67,064006;%%
  \bibitem{Bawin:2003dm}
  M.~Bawin and S.~A.~Coon,
  %``The singular inverse square potential, limit cycles and self-adjoint
  %extensions,''
  Phys.\ Rev.\  A {\bf 67}, 042712 (2003).
  %[arXiv:quant-ph/0302199].
  %%CITATION = PHRVA,A67,042712;%%
\bibitem{Camblong}
  H.~E.~Camblong and C.~R.~Ordonez,
  %``Anomaly in conformal quantum mechanics: From molecular physics to black
  %holes,''
  Phys.\ Rev.\  D {\bf 68}, 125013 (2003).
  %%CITATION = PHRVA,D68,125013;%%
  \bibitem{Ho}
  E.~J.~Mueller and T.~L.~Ho,
  arXiv:cond-mat/0403283.
  \bibitem{Braaten}
  E.~Braaten and D.~Phillips,
  %``The Renormalization group limit cycle for the 1/r**2 potential,''
  Phys.\ Rev.\  A {\bf 70}, 052111 (2004).
  %[arXiv:hep-th/0403168].
  %%CITATION = PHRVA,A70,052111;%%
  \bibitem{Barford:2004fz}
  T.~Barford and M.~C.~Birse,
  %``Effective theories of scattering with an attractive inverse-square
  %potential and the three-body problem,''
  J.\ Phys.\ A  {\bf 38}, 697 (2005).
  %[arXiv:nucl-th/0406008].
  %%CITATION = JPAGB,A38,697;%%
  \bibitem{Hammer:2005sa}
  H.~W.~Hammer and B.~G.~Swingle,
  %``On the limit cycle for the 1/r^2 potential in momentum space,''
  Annals Phys.\  {\bf 321}, 306 (2006).
  %[arXiv:quant-ph/0503074].
  %%CITATION = APNYA,321,306;%%
\bibitem{Hammer:2008ra}
  H.~W.~Hammer and R.~Higa,
  %``Discrete Scale Invariance and Long-Range Interactions,''
  Eur.\ Phys.\ J.\  A {\bf 37}, 193 (2008).
  %[arXiv:0804.4643 [nucl-th]].
  %%CITATION = EPHJA,A37,193;%%
\bibitem{FLS}
  W.~M.~Frank, D.~J.~Land and R.~M.~Spector,
  Phys.\ Rep. {\bf 43}, 36 (1971).
\bibitem{Calogero}
  F.~Calogero,
  %``Solution Of A Three-Body Problem In One-Dimension,''
  J.\ Math.\ Phys.\  {\bf 10}, 2191 (1969).
  %%CITATION = JMAPA,10,2191;%%
 \bibitem{Freedman}
  E.~D'Hoker and D.~Z.~Freedman,
  %``Supersymmetric gauge theories and the AdS/CFT correspondence,''
  arXiv:hep-th/0201253.
  %%CITATION = HEP-TH/0201253;%%
\bibitem{Sch}
C.~R.~Hagen,
%``Scale and conformal transformations in Galilean-covariant field theory,''
Phys.\ Rev.\  D {\bf 5}, 377 (1972);
%CITATION = PHRVA,D5,377;%%
U.~Niederer, 
%``The maximal kinematical invariance group of the free Schr\"odinger 
% equation,'' 
Helv.\ Phys.\ Acta {\bf 45}, 802 (1972).
\bibitem{Horava}
  P.~Horava and C.~M.~Melby-Thompson,
  %``Anisotropic Conformal Infinity,''
  arXiv:0909.3841 [hep-th].
  %%CITATION = ARXIV:0909.3841;%%
\bibitem{Blau}
  M.~Blau, J.~Hartong and B.~Rollier,
  %``Geometry of Schroedinger Space-Times, Global Coordinates, and Harmonic
  %Trapping,''
  JHEP {\bf 0907}, 027 (2009).
  %[arXiv:0904.3304 [hep-th]].
\bibitem{Ghosh}
  P.~K.~Ghosh and K.~S.~Gupta,
  %``On the Real Spectra of Calogero Model with Complex Coupling,''
  Phys.\ Lett.\  A {\bf 323}, 29 (2004).
  %[arXiv:hep-th/0310276].
  %%CITATION = PHLTA,A323,29;%%
\bibitem{Gubser}
  S.~S.~Gubser and I.~R.~Klebanov,
  %``A universal result on central charges in the presence of double-trace
  %deformations,''
  Nucl.\ Phys.\  B {\bf 656}, 23 (2003).
  %[arXiv:hep-th/0212138].
  %%CITATION = NUPHA,B656,23;%%
\bibitem{Witten}
  E.~Witten,
  %``Multi-trace operators, boundary conditions, and AdS/CFT correspondence,''
  arXiv:hep-th/0112258.
  %%CITATION = HEP-TH/0112258;%%
\bibitem{MS}
  S.~Moroz and R.~Schmidt,
  %``Nonrelativistic inverse square potential, scale anomaly, and complex
  %extension,''
  Annals Phys.\  {\bf 325}, 491 (2010).
  %[arXiv:0909.3477 [hep-th]].
  %%CITATION = APNYA,325,491;%%


\bibitem{Freedman98}
  D.~Z.~Freedman, S.~D.~Mathur, A.~Matusis and L.~Rastelli,
  %``Correlation functions in the CFT($d$)/AdS($d+1$) correspondence,''
  Nucl.\ Phys.\  B {\bf 546}, 96 (1999).
  %[arXiv:hep-th/9804058].
  %%CITATION = NUPHA,B546,96;%%

\bibitem{BH}
    E. Braaten, H.W. Hammer,
    Phys. Rept. \textbf{428}, 259 (2006).
    % eprint    =% "cond-mat/0410417",
\bibitem{Efimov}
  V.~Efimov, 
  Phys. Lett.  \textbf{33B}, 563 (1970);
  V.~Efimov, 
  Nucl. Phys. A  \textbf{210}, 157 (1973).
\bibitem{Kraemer}
  T.~Kraemer et al.,
  %``A universal result on central charges in the presence of double-trace
  %deformations,''
  Nature {\bf 440}, 315 (2006).
\bibitem{Nishida1}
  Y.~Nishida,
  private communication.
\bibitem{Fuertes}
C.~A.~Fuertes and S.~Moroz,
   %``Correlation functions in the non-relativistic AdS/CFT correspondence,''
   Phys.\ Rev.\  D {\bf 79}, 106004 (2009).
   %[arXiv:0903.1844 [hep-th]];%   %%CITATION = PHRVA,D79,106004;%%
\bibitem{Volovich}
 A.~Volovich and C.~Wen,
   %``Correlation Functions in Non-Relativistic Holography,''
   JHEP {\bf 0905}, 087 (2009).
   %[arXiv:0903.2455 [hep-th]].
   %%CITATION = JHEPA,0905,087;%%




\end{thebibliography}
\end{document}